# Electroluminescence from nitrogen-vacancy and interstitial-related centers in bulk diamond stimulated by ion-beam-fabricated sub-superficial graphitic micro-electrodes

J. Forneris[*,1,2], S. Ditalia Tchernij[1], A. Battiato[1,2], F. Picollo[2,1], A. Tengattini[1,2], V. Grilj[3], N. Skukan[3], G. Amato[4], L. Boarino[4], I. P. Degiovanni[4], E. Enrico[4], P. Traina[4], M. Jakšić[3], M. Genovese[4,2], P. Olivero[1,2]

[1] *Physics Department and NIS Inter-departmental Centre - University of Torino; via P. Giuria 1, 10125 Torino, Italy*
[2] *INFN - sez. Torino; CNISM - sez. Torino; via P. Giuria 1, 10125 Torino, Italy*
[3] *Ruđer Bošković Institute, Bijenicka 54, P.O. Box 180, 10002 Zagreb, Croatia*
[4] *Istituto Nazionale di Ricerca Metrologica (INRiM), Torino, Italy*

[*]jacopo.forneris@unito.it

**Abstract**
We report on the fabrication and characterization of a single-crystal diamond device for the electrical stimulation of light emission from nitrogen-vacancy ($NV^0$) and other defect-related centers. Pairs of sub-superficial graphitic micro-electrodes embedded in insulating diamond were fabricated by a 6 MeV $C^{3+}$ micro-beam irradiation followed by thermal annealing. A photoluminescence (PL) characterization evidenced a low radiation damage concentration in the inter-electrode gap region, which did not significantly affect the PL features dominated by NV centers. The operation of the device in electroluminescence (EL) regime was investigated by applying a bias voltage at the graphitic electrodes, resulting in the injection of a high excitation current above a threshold voltage (~300V), which effectively stimulated an intense EL emission from $NV^0$ centers. In addition, we report on the new observation of two additional sharp EL emission lines (at 563 nm and 580 nm) related to interstitial defects formed during MeV ion beam fabrication.

## 1. Introduction

Diamond is a promising material for several applications in quantum optics and photonics. In particular, its large energy band gap enables the access to a wide range of optically active centers with appealing properties for quantum cryptography, quantum computing and quantum sensing [2-6]. The negatively charged nitrogen-vacancy complex ($NV^-$) is arguably the most widely studied color center in diamond, due to its unique photo-physical properties (quantum efficiency, spin-sensitive transitions, high spin coherence time, etc.) at room temperature [7], while other appealing centers have been discovered and characterized in recent years. In general, most of these studies rely on the optical excitation of color centers, i.e. on their stimulation by means of laser light. The electrical stimulation of color centers in diamond would provide a practical tool for several specific applications, allowing to simply integrate these appealing quantum systems into compact devices. The observation of electroluminescence (EL) from $NV^0$ centers has been reported in previous works, both in the case of ensemble emission [8,9] and at the single-photon emitter level [10,11].

These studies typically relied on controlled doping to fabricate current-injecting p-i-n structures. Deep Ion Beam Lithography (DIBL) is an alternative strategy to the technologically challenging n-type diamond doping, allowing the injection of charge carriers in diamond through ion-beam-fabricated graphitic electrodes [12,13]. In particular, DIBL offers the advantage of allowing a flexible design in the definition of the electrodes geometry with micrometric resolution, particularly with regards to the possibility of creating sub-superficial structures. This latter feature represents a significant advantage for the electrical stimulation and control color centers located in depth in the bulk crystal, which are characterized by superior opto-physical properties due to the distance from surface defects [14].

The DIBL technique exploits the sharp Bragg's peak of MeV ions to selectively convert sub-superficial regions of single-crystal diamond to an amorphous carbon phase through the localized introduction of radiation-induced lattice damage. These critically damaged regions can be subsequently converted to a graphitic phase upon high-temperature thermal annealing [15,16]. Therefore, by means of DIBL it is possible to directly fabricate highly conductive graphitic micro-electrodes embedded in the insulating diamond matrix, which can be exploited to generate local electric fields and/or inject currents in the material bulk, with appealing applications in ionizing radiation detection [17], cellular bio-sensing [18], IR emission [19], etc. The DIBL technique was recently applied for the stimulation of EL in diamond in [12], where it was demonstrated that EL can be stimulated by means of a controlled current injection from DIBL-fabricated graphitic micro-electrodes. Besides $NV^0$ emission, the spectral analysis of the EL emission revealed features associated with the formation of extended defects (i.e. lattice dislocations) and with the implantation of foreign atoms (i.e. He) during the DIBL fabrication process.

The aim of this work is to demonstrate the fabrication of a new diamond-based EL source unaffected by background luminescent emission from foreign impurities (with the exception of the intentional N doping) or extended defects, as well as to study potentially appealing sharp EL emission lines from intrinsic point defects. With this purpose, a 6 MeV $C^{3+}$ ion beam was employed for the DIBL fabrication of graphitic micro-electrodes, in order to define an all-carbon-based device. Subsequently, the fabricated structure was characterized in its light emission properties, by means of both punctual and spatially resolved photoluminescence (PL) and EL measurements.

## 2. Experimental

*2.1 Device fabrication*

The employed sample is a 3×3×0.3 mm³ type IIa single-crystal CVD diamond produced by Element Six, with nominal substitutional nitrogen and boron concentrations of <1 ppm and <0.05 ppm, respectively. The crystal orientation is <100> and the sample was optically polished on both of the larger faces. Two graphitic electrodes were directly fabricated in the diamond bulk by raster-scanning a Ø ~10 μm focused 6 MeV $C^{3+}$ beam along linear paths using the 6 MV tandem accelerator of the Laboratory for Ion Beam Interactions of the Ruđer Bošković Institute. The ion penetration depth, estimated using SRIM Monte Carlo simulation code [20], was of ~3 μm below the sample surface (Fig. 1a). The ion fluence (~4×10¹⁶ cm⁻²) was set to introduce at the Bragg's peak a radiation-induced vacancy density above the threshold value (9×10²² cm⁻³ [17]) required to create ~350 nm thick amorphous channels.

The sample was subsequently annealed in vacuum for 2 hours at 1000 °C, to convert the aforementioned amorphous channels to a graphitic phase and to concurrently recover the radiation damage in the regions damaged at a lower density. After the thermal treatment, the sample underwent a surface cleaning process in oxygen plasma (30 min, 2.5×10⁻² mbar pressure, 30 W RF power, 20 sccm $O_2$ flux) to remove surface conductivity from graphitic surface defects formed during the high-temperature annealing process. A focused ion beam (FIB) milling with 30 keV $Ga^+$ ions (FEI Quanta 3D™ Dualbeam) was performed at the external endpoints of the graphitic channels (see Fig. 1c), followed by a 70 nm Ag deposition through a patterned mask, to define electrical contacts, which were wire-bonded to provide a connection with an external voltage source. An optical micrograph of the micro-fabricated structure is shown in Fig. 1b. The resulting device consists of two independent ~10 μm wide, ~100 μm long electrodes fabricated along the same axis and spaced by a distance *d* ~9 μm.

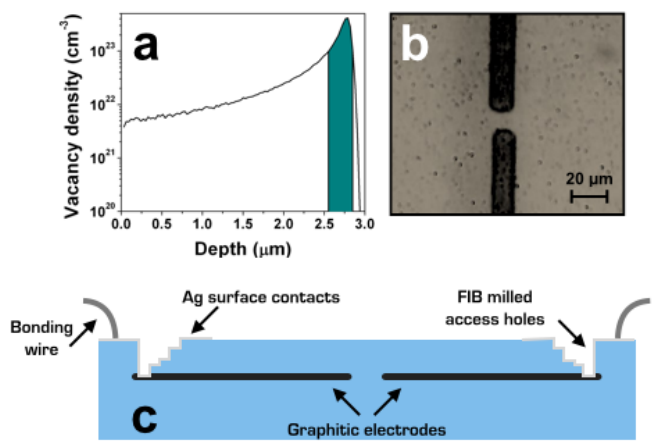

**Fig. 1:** **a)** SRIM simulation of the vacancy density profile associated with the implantation of 6 MeV $C^{3+}$ ions at a fluence of 4×10¹⁶ cm⁻². The region highlighted in green indicates the depth at which the vacancy density exceeds the graphitization threshold. **b)** Optical micrograph of the fabricated structure with graphitic electrodes. The arrows indicate the regions of the device investigated with PL and EL measurements. **c)** Schematics of the device cross-section. Two graphitic electrodes are exposed to the surface at their endpoints through FIB milling. A 60 nm Ag deposition ensures their electrical connection with an external voltage supply.

*2.2 Photoluminescence*

Spectral PL and EL measurements were performed on the device to study its suitability to electrically stimulate the $NV^0$ centers, as well as to assess the ion irradiation effects on its light emission properties. PL spectra were acquired under 532 nm laser excitation using a Jobin Yvon Raman micro-spectrometer equipped with a CCD Andor "DU420A-OE" detector. The spatial resolution of the adopted 20× air objective was ~5 μm, allowing to probe the spectral features of both the graphitic electrodes and of their spacing gap region. The focal depth was comparable with the Bragg's peak of 6 MeV $C^{3+}$ ions in diamond, thus providing insufficient spatial resolution to unequivocally discriminate the luminescence arising from the buried graphitic regions and the overlying damaged diamond cap layer.

In Fig. 2a a typical PL spectrum acquired from the inter-electrode gap (area labeled as "1" in Fig. 2c) is shown. The spectrum exhibits the typical features of an "optical grade" diamond sample, such as the intense first-order Raman peak at 572 nm (corresponding to 1332 cm⁻¹ Raman shift), and the $NV^0$ and $NV^-$ zero phonon lines (ZPLs) respectively at 575 nm and 638 nm, together with their associated phonon sidebands (PS) [8]. Additionally, a weak peak at 740 nm is attributed to residual GR1 centers, i.e. to isolated vacancies created by the im-plantation of stray ions in the inter-electrode gap region and not completely annealed out upon the thermal treatment [12]. It is worth stressing that, as expected from the implantation of carbon ions, no centers associated with the implantation of foreign atoms (other than the native N atoms) were observed.

In Fig. 2b a typical PL spectrum acquired from a graphitic electrode (area labeled as "2" in Fig. 2c) is clearly affected by the significant structural modifications introduced by the ion fabrication technique. The spectrum displays a significantly attenuated Raman peak, indicating the highly distorted structure of the implanted crystal and a lower concentration of residual $sp_3$ bonds in the lattice, as an expected consequence of the conversion of the buried channel to a graphitic structure. Moreover, the ZPLs of the NV centers are not visible. The PL spectrum is dominated by a large emission in the 740-800 nm range, which is typically associated with the formation of the radiation B-band [11,21] associated to lattice dislocations. The whole spectrum shows a periodic intensity modulation which is attributed to interference fringes caused by mutiple internal reflections of the stimulated background luminescence occurring between the sample surface and the buried graphitic layers [22].

A PL map of the device (Fig. 2c) was acquired using a home-built confocal microscopy setup equipped with a single-photon sensitive avalanche detector. PL emission was stimulated with a 633 nm laser excitation (0.85 mW power) and detected employing a 730 nm long-pass filter to highlight the spatial distribution of the aforementioned B-band spectral component. The features of the PL map highlight the shape of the graphitic electrodes, where a high photon count rate is measured. Low (<10 kcps) photon count rates are observed outside the ion-irradiated areas, confirming that the radiation-related defects are confined to the graphitic region. This result indicates that the amount of structural damage introduced in the inter-electrode gap region does not prevent the exploitation of the device to electrically stimulate isolated color centers.

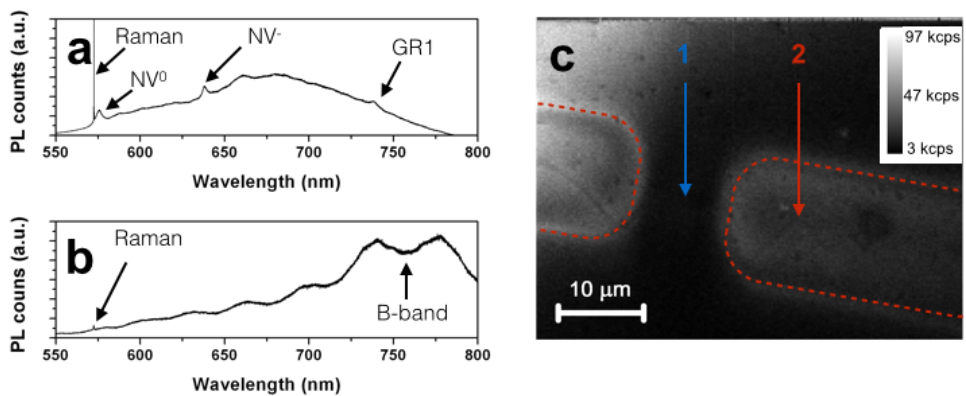

**Fig. 2: a)** PL spectrum acquired from the inter-electrode gap region labelled as "1" in Fig. 2c. **b)** PL spectrum acquired from the region in correspondence of one of the two buried electrodes, labelled as "2" in Fig. 2c. **c)** Confocal PL map acquired with a 730 nm long pass filter from the region including the edges of the graphitic electrodes and their spacing gap region. The higher photon counts in correspondence of the electrodes indicate the presence of radiation-induced defects (B-band). The electrodes edges are highlighted by the red dashed lines.

*2.3 Electroluminescence*

An electrical characterization of the device was performed by acquiring its current-voltage (*I-V*) characteristics (voltage step: 1V; source time: 0.3 s), as shown in Fig. 3a. The I-V curve exhibits a linear behavior in the 0-300 V range, with a typical current of ~250 nA at 300 V. When the bias voltage is further increased, the device switches to a high-current regime, reaching values of 0.1 mA at an applied bias of 500 V. The electrical response of the device at >300 V bias was stable over time and enabled the investigation of induced EL, which was detectable for currents higher than 10 μA. An optical micrograph acquired at a bias voltage of 500 V is shown in Fig. 3b, exhibiting an intense red emission visible by naked eye in the region labelled as "1" in Fig. 2c. EL spectra acquired at 350 V and 500 V bias with the same experimental setup adopted for PL measurements are reported in in Fig. 3c and Fig. 3d, respectively. The main spectral feature is the emission from the $NV^0$ center, which is clearly characterized by the ZPL at 575 nm and its PS at higher wavelengths. The $NV^-$ component is absent, consistently with previous observations indicating that the $NV^-$ center is not active under electrical excitation [8,10-13,21]. Moreover, two sharp EL emission lines (not active under optical excitation at 532 nm), are clearly visible at 563 nm and 580 nm wavelengths.

A comparison of the relative intensity of the aforementioned emission lines (Fig. 3c and Fig. 3d) suggests a correlation of the two ZPLs, which can be tentatively attributed to the same defect family or to different charge states of the same complex. These previously unreported sharp EL emission lines are attributed to self-interstitial defects, i.e. pairs of nearby iterstitial atoms along the <100> axis, previously observed in cathodoluminescence (CL) and PL ($\lambda_{exc}$ = 488 nm) measurements performed on ion- and electron-irradiated diamonds [21,23].

While these observations indicate that the aforementioned sharp EL lines could origin from the implantation of stray ions in the inter-electrode gap, the absence of significant shifts and modifications to the first-order Raman peak shown in Fig. 2a suggests that the 563 nm and 580 nm ZPLs are not associated with extended defects or platelets [24,25], and could be thus be ascribed to EL from point defects. Finally, it is worth noting that, differently from what reported in previous works on sub-superficial graphitic electrodes, no spectral features associated with either the B-band [11], the GR1 peak [12] or the A-band (a broad emission centered at 435nm [12,21]) were observed from the active inter-electrode region of the device under investigation, indicating the overall quality of the crystal lattice.

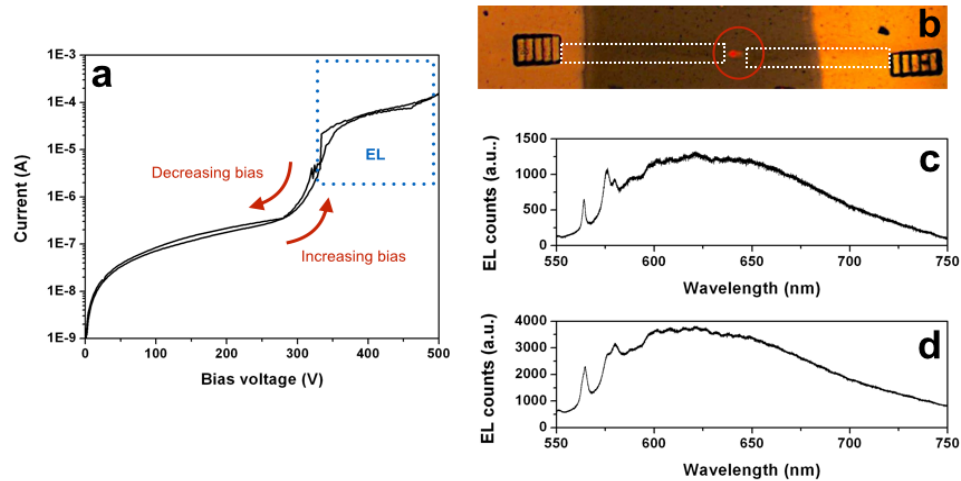

**Fig. 3: a)** Current-voltage characteristics of the device . The dashed blue square indicates the regime in which EL was observed. **b)** Optical micrograph of the device biased at a voltage of 500 V. The shape of the graphitic electrodes is indicated by the white dashed line as an eye guide. The intense EL from NV centers is clearly visible as the red spot in the inter-electrode gap region (circled in red). **c)** EL spectrum acquired at a bias voltage of 350 V from the inter-electrode gap (red circle in Fig. 3b and area labelled as "1" in Fig. 2c). **d)** EL spectrum acquired at the same position at an applied bias of 500V.

**Conclusions**

In this work, we reported on the DIBL fabrication of an all-carbon device for the electrical stimulation of color centers in diamond, consisting of coupled buried graphitic electrodes spaced by a distance of 9 μm. A PL analysis of the graphitic electrodes enabled to qualify the structural properties of the heavily irradiated regions of the fabricated structure, as well as to investigate the possible presence of residual radiation damage in the active region of the device and to qualify its effects on the EL emission. The PL characterization of the device evidenced the light emission from the highly defective region overlying the sub-superficial graphitic electrodes, dominated by the B-band contribution. A spatial mapping of the B-band emission was performed using a confocal microscopy setup, revealing a negligible defect concentration in the active region of the device, i.e. in the inter-electrodes gap.

An intense EL dominated by $NV^0$ emission was stimulated by current injection at operating voltages between 300 V and 500 V, for which a significant increase of injected current was measured. Differently from devices fabricated by $He^+$ ion irradiation [12], the implantation of $C^{3+}$ ions prevented the formation of additional color centers related to foreign impurities in the diamond lattice. Moreover, in the EL spectra no emission associated to dislocation- and vacancy-related defects (A-band, GR1 center, B-band) was observed, indicating that the ion irradiation technique was effective in the fabrication of the buried electrodes without introducing a significant amount of radiation damage.

Interestingly, previously unreported sharp EL emission peaks at 563 nm and 580 nm were observed and tentatively attributed to interstitial defects, consistently with previous CL studies. The intensity of these peaks exhibits a stronger dependence on the injection current with respect to the $NV^0$ center, indicating an efficient EL process which could represent an appealing candidate for the development of electrically stimulated single-photon sources from individual defects, a tool of the utmost importance for emerging quantum technologies.

The results presented in this work indicate a possible development of solid-state devices based on the electrical stimulation NV centers in diamond, which could be achieved by a significant reduction in the operating voltage and the prevention of the introduction of any radiation damage in the inter-electrodes gap, i.e. through the exploitation of the DIBL technique on patterned implantation masks with sub-micrometric resolution [26]. In perspective, the deterministic implantation [27], the fabrication by means of femtosecond laser pulse [28] or the incorporation during CVD growth [29] of NV centers or foreign atoms would lead to the formation of deep EL-active centers that could be electrically excited and/or controlled by sub-superficial graphitic electrodes in a stable bulk environment. The depth of such electrodes can be easily defined by tuning the energy and species of implanted ions in the DIBL process, thus allowing an ideal matching between the respective depths. Finally, the technique would allow in perspective to define arrays and structures of electrically-controlled emitters in the diamond bulk, which are characterized by superior opto-physical properties due to the distance from surface defects [14].

**Acknowledgements**

This research activity was funded by the following projects, which are gratefully acknowledged: FIRB 'Future in Research 2010' project (CUP code: D11J11000450001) funded by the Italian Ministry for Teaching, University and Research (MIUR); EMRP project 'EXL02-SIQUTE' (jointly funded by the EMRP participating countries within EURAMET and the European Union); EMPIR project 14IND05-MIQC2 (co-funded by the European Union's Horizon 2020 research and innovation programme and the EMPIR Participating States); the 'A.Di.N-Tech.' project (CUP code: D15E13000130003) funded by the University of Torino and Compagnia di San Paolo in the framework of the 'Progetti di ricerca di Ateneo 2012' scheme and in the support of Nanofacility INRiM.